\documentclass[twocolumn,aps,prl,preprintnumbers,superscriptaddress]{revtex4-2}
\usepackage[utf8]{inputenc}
\usepackage{mathrsfs}
\usepackage{amsmath}
\usepackage{amsfonts}
\usepackage{amssymb}
\usepackage{amsthm}
\usepackage{multirow}
\usepackage{color}
\usepackage{graphicx}
\usepackage{pifont}
\usepackage{lettrine}
\graphicspath{{./images/}}
\usepackage{geometry}
\usepackage{hyperref}
\usepackage{lineno}
\usepackage{textcomp}

\hypersetup{
     unicode=false,          
     pdftoolbar=true,        
     pdfmenubar=true,        
     pdffitwindow=false,     
     pdfstartview={FitH},    
     pdftitle={My title},    
     pdfauthor={Author},     
     pdfsubject={Subject},   
     pdfcreator={Creator},   
     pdfproducer={Producer}, 
     pdfkeywords={keyword1} {key2} {key3}, 
     pdfnewwindow=true,      
     colorlinks=true,       
     linkcolor=blue,     
     citecolor=blue,        
     filecolor=blue,      
     urlcolor=blue           
}
\usepackage[toc,page]{appendix}
\usepackage{ulem}

\begin{document}
\makeatletter
\@ifundefined{textcolor}{}
{
\definecolor{BLACK}{gray}{0}
\definecolor{WHITE}{gray}{1}
\definecolor{RED}{rgb}{1,0,0}
\definecolor{GRAY}{rgb}{0.4,0.4,0.4}
\definecolor{GREEN}{rgb}{0,1,0}
\definecolor{BLUE}{rgb}{0,0,1}
\definecolor{CYAN}{cmyk}{1,0,0,0}
\definecolor{MAGENTA}{cmyk}{0,1,0,0}
\definecolor{YELLOW}{cmyk}{0,0,1,0}
}
\makeatother

\title{Ferron-driven motion of non-collinear ferroelectric domain walls}
\author{Huanhuan Yang}
\email{yang.huanhuan.b6@tohoku.ac.jp}
\affiliation{WPI-AIMR, Tohoku University, 2-1-1 Katahira, Sendai 980-8577, Japan}
\affiliation{Center for Spintronics Research Network, Tohoku University, Sendai 980-8577, Japan}
\author{Gerrit E. W. Bauer}
\affiliation{WPI-AIMR, Tohoku University, 2-1-1 Katahira, Sendai 980-8577, Japan}
\affiliation{Center for Spintronics Research Network, Tohoku University, Sendai 980-8577, Japan}
\affiliation{Institute for Materials Research, Tohoku University, 2-1-1 Katahira, Sendai 980-8577, Japan}

\begin{abstract}
We study the propagation of ferrons, the quanta of electric polarization waves, through a ferroelectric texture. We predict by analytic and numerical calculations that a $90^\circ$ domain wall (DW) operates as a beam splitter of longitudinal ferron and transverse optical phonon modes. An associated momentum-flux imbalance leads to forces that drive DW motion. In the weak-anisotropy limit, the DW becomes a rotating texture with nearly fixed-modulus, whose low-energy dynamics is described by a sine-Gordon equation. A ferron source then attracts the DW through a damping-induced force. Both driving forces scale linearly with incident wave intensity. Our results uncover qualitatively different routes to driven ferroelectric DW motion enabled by the vector nature of the polarization and offer functionalities for THz ferroelectric devices.
\end{abstract}

\maketitle  

{\color{blue} {\it Introduction.}}
Domain walls (DWs) are ubiquitous interfaces between distinct ordered states, such as those in magnetic and ferroelectric materials, offering promising functionalities for nanoelectronics \cite{Catalan2012}. Ferroelectric DWs are often Ising-like, with a collinear polarization that vanishes in its center. Advances in materials engineering and characterization revealed non-Ising DWs with multicomponent polarization textures \cite{CherifiHertel2017,Nataf2020,Mantri2021}, including Bloch-like \cite{He2024ActaMat} and N\'eel-like \cite{Wei2016} walls in which the polarization rotates with a nearly constant modulus.  $90^\circ$ ferroelectric DWs are common in tetragonal perovskites \cite{Cao1991,Hlinka2006,Morikawa2023,Matzen2014,Tang2014,Karthik2012,Hirohashi2003}, hafnia-based thin films \cite{Zheng2023,Wen2026}, and 2D group-IV monochalcogenides \cite{Fei2016,BarrazaLopez2021,Yan2022} with polarization textures that vary in both modulus and direction.
Unconventional DW textures may facilitate DW motion and polarization switching \cite{Gao2013}. Studies of their dynamics lead to fundamental insights into non-equilibrium and non-linear phenomena \cite{Shin2007,Jo2009,Kagawa2016} and promise a superior ferroelectric nanoelectronics \cite{Sharma2019,Meier2022,Junquera2023,McCluskey2025}.

Magnetic memory technologies routinely store information in metallic magnets by moving DWs and switch magnetizations by spin-polarized electric currents \cite{Yamaguchi2004,Parkin2008}. In magnetic insulators, spin waves or magnons can do the same \cite{Yan2011,Wang2015DMI,Kim2015,Lan2022,Hinzke2011,Kim2012,Schlickeiser2014,Wang2012Emission,Kim2019Unidirectional}.
Stimulated by those insights and the prospect of ultrafast THz polarization dynamics, we recently established that a flux of ferrons, the quanta of polarization waves in the ferroelectric order \cite{Bauer2021,Tang2022,Tang2024Ferron,Adachi2023,RodriguezSuarez2024,Choe2026,Shen2025,Zhang2025,Subedi2026,Manna2026} can move ferroelectric Ising DWs \cite{Yang2026}. However, since the polarization transfer force kicks in only to second order in the applied current, a more power-efficient mechanism is desirable.

In this Letter, we derive the interaction of polarization waves with $90^\circ$ ferroelectric DWs and identify two distinct driving mechanisms, both linear in the incident wave intensity. In a monodomain of a simple displacive ferroelectric, the optical phonons along the equilibrium polarization oscillate in a strongly non-parabolic potential which endows them with a dc electric dipole moment \cite{Tang2022}, while the modes polarized transverse to the order are conventional optical phonons. A non-collinear DW mixes these modes and enables coherent ferron-phonon conversion. In certain cases, mode conversion makes the incoming and outgoing momentum fluxes unequal, resulting in a DW force whose direction depends on the incident wave polarization. In the limit of weak crystal anisotropy, the equilibrium DW is N\'eel-like with an adiabatically rotating constant-modulus polarization. Damping-induced attenuation generates a spatial gradient of the polarization-wave intensity, which couples to the DW texture and drives the wall toward the source. Our results suggest strategies for using ferroelectric DWs to mix or split phonon and ferron modes and control DW motion in ferroelectric devices.

\begin{figure}
\centering
\includegraphics[width=0.49\textwidth]{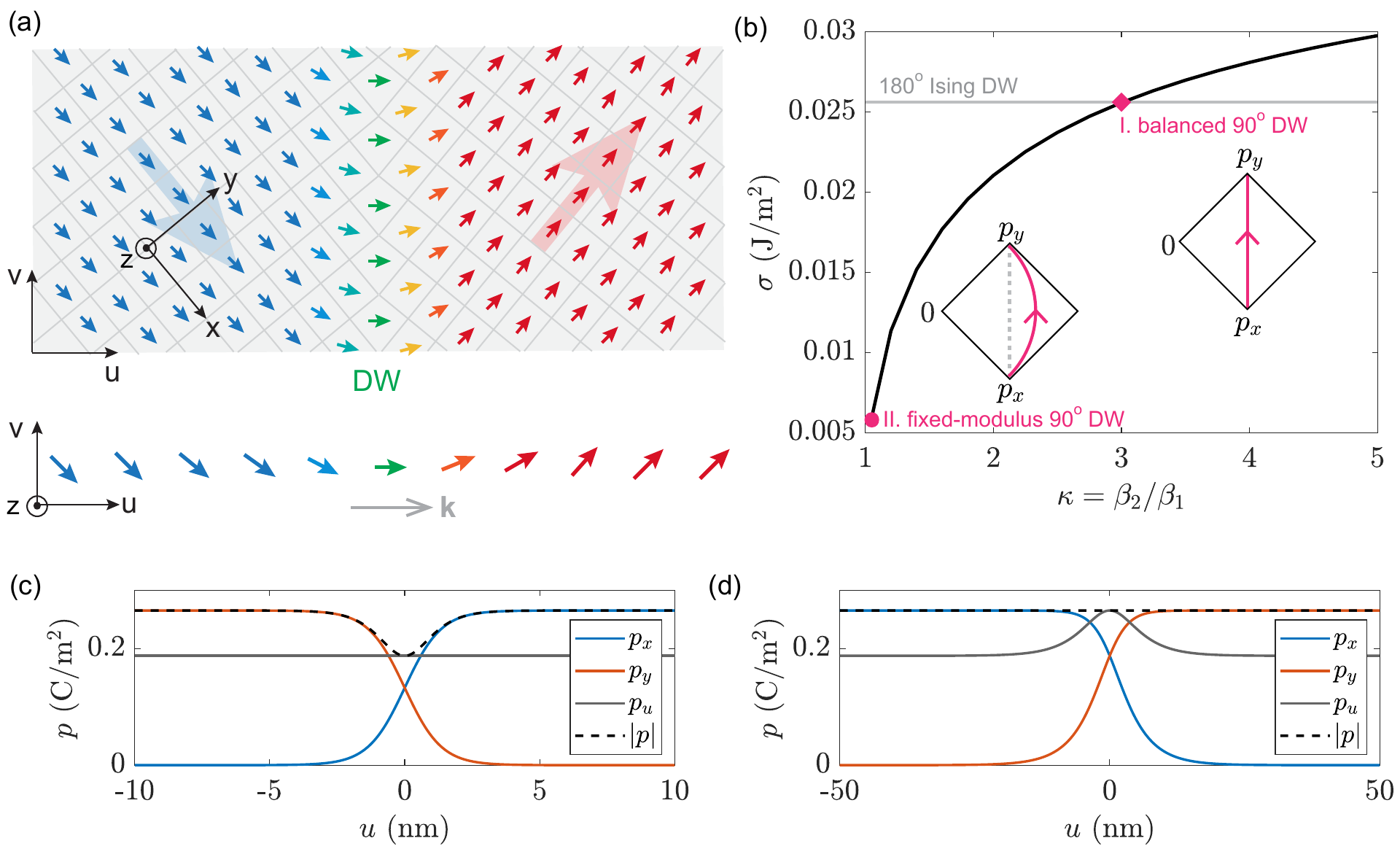}
\caption{(a) Illustration of a $90^\circ$ head-to-tail  DW in a ferroelectric film (top) and the corresponding effective 1D model along $\hat{u}$ (bottom). (b) DW energy $\sigma$ versus anisotropy ratio $\kappa=\beta_2/\beta_1$. Gray line: $180^\circ$ Ising-wall energy. Black curves: energy of two $90^\circ$-DWs. Insets:   $p_x(u)$ \textit{vs}. $p_y(u)$ textures normal to the DW. (c) Polarization profile of a balanced $90^\circ$ DW for $\kappa=3$  with $p_x+p_y=p_s$. (d) Fixed-modulus $90^\circ$ DW for $\kappa=1.05$ ($|p|=p_s$).}\label{F1}
\end{figure}

{\color{blue} {\it DW profiles and ferron modes.}}
The Ginzburg-Landau-Devonshire free energy of a ferroelectric thin film with in-plane polarization ${\bf p}=(p_x,p_y)$ reads \cite{Cao1991,Hlinka2006,Tagantsev2009,Ishibashi2002,Azuma2023}
\begin{equation}\label{Free_energy}
\begin{aligned}
{\mathcal F}=&\int dxdy~f_0+\frac{1}{2}\alpha\,(p_x^2+p_y^2)+\frac{1}{4}\beta_1\,(p_x^4+p_y^4)\\
&+\frac{1}{2}\beta_2\,p_x^2p_y^2 + g\left[(\partial_x p_x)^2+(\partial_y p_y)^2\right]-{\bf E}\cdot{\bf p},
\end{aligned}
\end{equation}
where  $\alpha<0$ below the critical temperature, $\beta_{1,2}>0$, $g$ parameterizes the stiffness, and ${\bf E}$ is an applied electric field.

We consider the 90$^\circ$ head-to-tail DW sketched in Fig.~\ref{F1}(a) that connects the two in-plane domains with $(p_x,p_y)=(p_s,0)$ and $(0,p_s)$, where $p_s=\sqrt{|\alpha|/\beta_1}$. The unit vectors $\hat u=(\hat x+\hat y)/\sqrt{2}$ and $\hat v=(\hat y-\hat x)/\sqrt{2}$ are the normal and tangent to the DW, respectively, and span a rotated coordinate system in which  $u=\frac{\sqrt{2}}{2}(x+y)$, $v=\frac{\sqrt{2}}{2}(y-x)$, $p_u=\frac{\sqrt{2}}{2}(p_x+p_y)$, and $p_v=\frac{\sqrt{2}}{2}(p_y-p_x)$. Translational invariance along $\hat v$ implies that the DW ground state energy depends only on $u$ and $p_{u,v}(u)$. The free energy per unit length becomes
\begin{equation}\label{Free_energy_1D}
\begin{aligned}
&{\mathcal F}_{1{\rm D}}
= \int du~
 f_0+\frac{\alpha}{2}(p_u^2+p_v^2)
+\frac{\beta_1+\beta_2}{8}(p_u^4+p_v^4)  \\
&+\frac{3\beta_1-\beta_2}{4}p_u^2p_v^2
+\frac{g}{2}[(\partial_u p_u)^2+(\partial_u p_v)^2]
-{\bf E}\cdot{\bf p}.
\end{aligned}
\end{equation}
The polarization dynamics obeys the Landau-Khalatnikov-Tani (LKT) equations \cite{Tani1969,Sivasubramanian2004}
\begin{equation}\label{LKT}
\begin{aligned}
&m_p\ddot p_u+\gamma\dot p_u
-g\partial_u^2 p_u\\
&+\alpha p_u
+\frac{1}{2}(\beta_1+\beta_2)p_u^3
+\frac{1}{2}(3\beta_1-\beta_2)p_v^2 p_u 
=E_u,\\
&m_p\ddot p_v+\gamma\dot p_v
-g\partial_u^2 p_v\\
&+\alpha p_v
+\frac{1}{2}(\beta_1+\beta_2)p_v^3
+\frac{1}{2}(3\beta_1-\beta_2) p_u^2 p_v
=E_v,
\end{aligned}
\end{equation}
where $m_p$ is an inertial mass and $\gamma$ is a damping constant.

The ground state textures minimize the free energy or, equivalently, solve the LKT equations for $\ddot{\mathbf{p}}=\dot{\mathbf{p}}=0$. Both $180^\circ$ Ising and $90^\circ$ wall DWs are solutions, the former (latter) being stable when the anisotropy ratio $\kappa=\beta_2/\beta_1$ is larger (smaller) than 3.  We obtain analytic solutions for the $90^\circ$ DW in two limiting cases of the anisotropy ratio $\kappa=\beta_2/\beta_1$ that we refer to as ``balanced" and ``fixed modulus", see Fig.~\ref{F1}(b): 

I. Balanced wall: for $\kappa=3$, the polarizations read $p^0_u(u)=\frac{\sqrt{2}}{2}p_s$ and $p^0_v(u)=-\frac{\sqrt{2}}{2}p_s \tanh\frac{u-U}{\Delta_{\rm I}}$, where $\Delta_{\rm I}=\sqrt{2g/|\alpha|}$ is the DW width, and are plotted in Fig.~\ref{F1}(c).

II. Fixed-modulus wall: in the weak-anisotropy limit  $\kappa\to1^+$, the DW becomes a rotating texture with nearly constant modulus of the polarization. Writing $p_u=p_s\cos\theta$ and $p_v=p_s\sin\theta$, with $\theta$ measured from the $u$ axis, the equilibrium DW is described by $\theta_0(u)=-\frac{\pi}{4}+\arctan\left[\exp\left(\frac{u-U}{\Delta_{\rm II}}\right)\right]$ with $\Delta_{\rm II}=\frac{\sqrt{g}}{p_s\sqrt{\beta_2-\beta_1}}$. Figure \ref{F1}(d) shows that this wall is much wider and diverges for $\kappa=1$, which reduces the elastic energy cost without having to pay much anisotropy energy.

\begin{figure}
  \centering
  \includegraphics[width=0.48\textwidth]{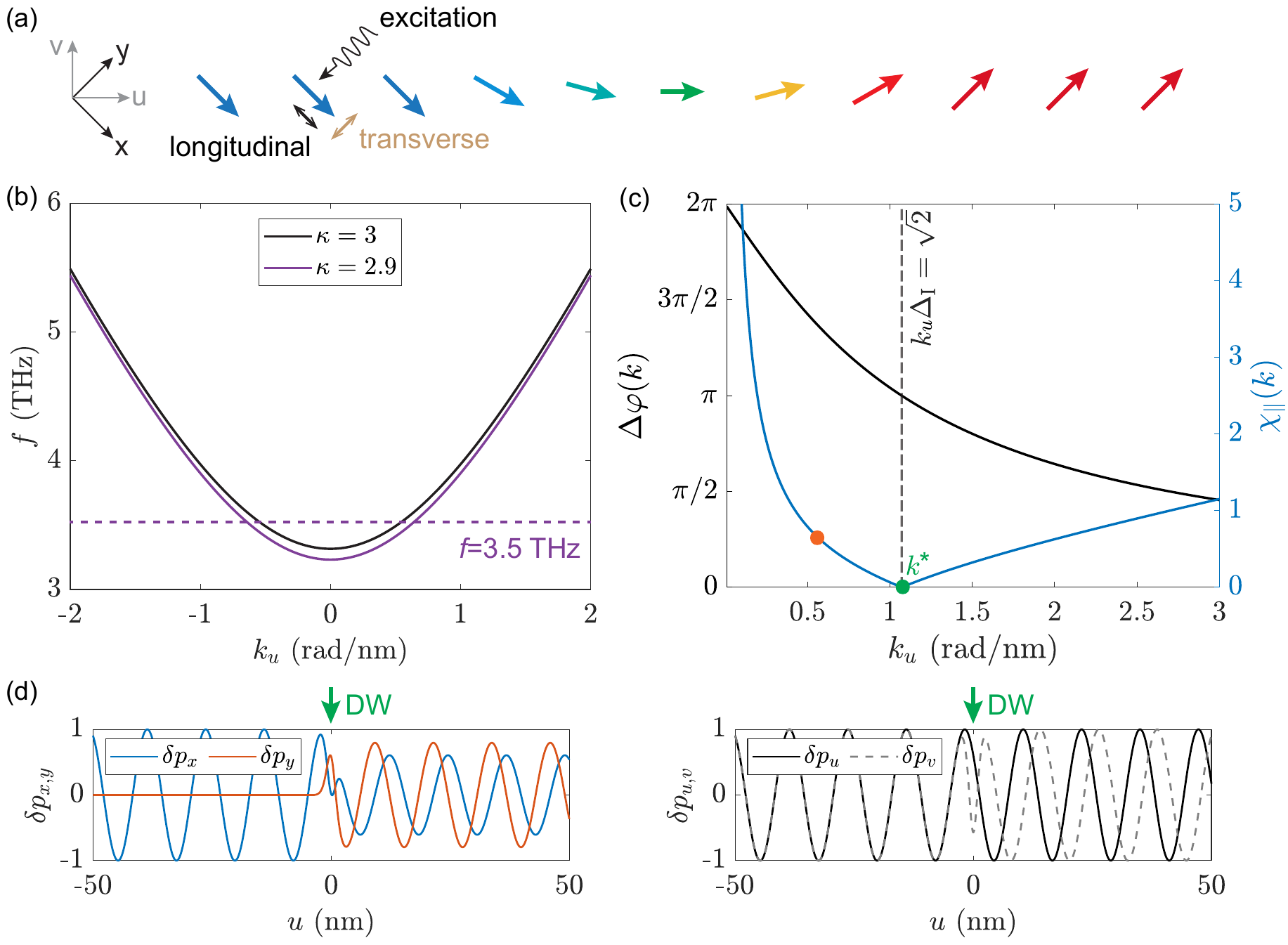}
  \caption{(a) $90^\circ$ DW with longitudinal and transverse excitations and laboratory \((u,v)\) as well as local \((x,y)\) coordinate systems. 
(b) Transverse optical phonon dispersions in a uniform domain for $\kappa=3$ (black) and $\kappa=2.9$ (purple). The longitudinal ferron branch (dashed line) does not depend on $\kappa$ and coincides with the transverse one for $\kappa=3.$
(c) Relative phase shift $\Delta\varphi(k)$ (black, left axis) and transmission amplitude ratio $\chi_{\parallel}(k)$ (blue, right axis) of an incident ferron (longitudinal mode). 
(d)  Normalized transmission amplitudes $\delta p_u$ and $\delta p_v$ of an incoming ferron at $f=3.5$ THz (left) and the amplitudes in the lattice system $\delta p_x$ and $\delta p_y$ (right).}\label{F2}
\end{figure}

Next, we focus on the in-plane polarization dynamics, because the out-of-plane fluctuations are pushed to high frequencies by the depolarizing field. In a uniform domain the longitudinal polarization-carrying ferrons $\delta p_{\parallel}=(\delta p_u-s\delta p_v)/\sqrt{2}$ parallel to the equilibrium ${\bf p}_0$  are orthogonal to the optical transverse phonons $\delta p_{\perp}=(\delta p_u+s\delta p_v)/\sqrt{2}$ perpendicular to ${\bf p}_0$, viz. $p_{\parallel}(u,t)=p_s+\delta p_{\parallel}(u,t)$ and $p_{\perp}(u,t)=\delta p_{\perp}(u,t)$, as depicted in Fig.~\ref{F2}(a). Here, \(s=+1\) and \(-1\) label the \(x\)- and \(y\)-polarized domains, respectively. Linearizing Eq.~\eqref{LKT} for \(\gamma=E=0\), the two bands have  frequency dispersions
\begin{equation}\label{Dispersion}
\begin{aligned}
m_p\omega_{\parallel}^2=gk_u^2+2|\alpha|,~~
m_p\omega_{\perp}^2=gk_u^2+|\alpha|(\kappa-1).
\end{aligned}
\end{equation}
The gap of the longitudinal amplitude mode is $\sqrt{2|\alpha|/m_p}$, while $\sqrt{|\alpha|(\kappa-1)/m_p}$ of the transverse mode is controlled by the anisotropy and vanishes for $\kappa= 1$. We plot the dispersions in Fig.~\ref{F2}(b) using the parameters of BaTiO$_3$ \cite{SM}.

{\color{blue} {\it I. Balanced wall.}}
Focusing on $\kappa=3$, we linearize Eq. \eqref{LKT} as $p_{u,v}(u,t)=p_{u,v}^0(u)+\delta p_{u,v}(u,t)$. The polarization in the $u$ channel is a plane wave in a uniform potential, whereas the $v$ channel mode feels a P\"oschl-Teller potential that supports two bound and otherwise reflection-less scattering states \cite{SM}. An arbitrary incident polarization wave can be decomposed into $u$ and $v$ components. Both transmit the DW without backscattering but they acquire the transmission phase shift 
\begin{equation}
\Delta\varphi= -2\arctan\left[\frac{3k_u\Delta_{\rm I}}{2-(k_u\Delta_{\rm I})^2}\right](\mathrm{mod}~2\pi)
\end{equation}
as the black curve in Fig.~\ref{F2}(c). When transforming back to the laboratory coordinates, \(\Delta\varphi \ne 0\)  implies a beam splitting functionality of the DW. The transmission amplitude of a longitudinal mode $\delta p_{\parallel}$ incident from the left has both $\delta p_{\parallel}$ and $\delta p_{\perp}$ components, as illustrated in Fig.~\ref{F2}(d). The transmitted amplitude ratio is
\begin{equation}\label{chi}
\chi_{\parallel}(k_u)
=\frac{\delta p_\perp}{\delta p_\parallel}
=i\cot\frac{\Delta\varphi}{2}
=i\frac{(k_u\Delta_{\rm I})^2-2}
{3k_u\Delta_{\rm I}} .
\end{equation}
Since $\chi_{\parallel}(k_u)$ is purely imaginary, the transmitted longitudinal and transverse components have a relative phase of $\pm\pi/2$, while its modulus $|\chi_{\parallel}(k_u)|$ is plotted as the blue curve in Fig.~\ref{F2}(c). The transmitted wave remains purely longitudinal only at $k^*=\sqrt{2}/\Delta_{\rm I}$. For an incident transverse mode, $\chi_{\perp}(k)=\chi_{\parallel}^{-1}(k)$. Since  \(\omega_{\parallel}=\omega_{\perp}\) for  $\kappa=3$ the hybridization of the modes conserves an incoming the linear momentum-flux that therefore does not exert a force on the DW. We cannot solve the problem analytically for  $\kappa \ne 3$ but we anticipate that non-degeneracy of ferrons and transverse optical phonons leads to a force on the DW, as shown in the following by numerical simulations.

\begin{figure}
  \centering 
  \includegraphics[width=0.49\textwidth]{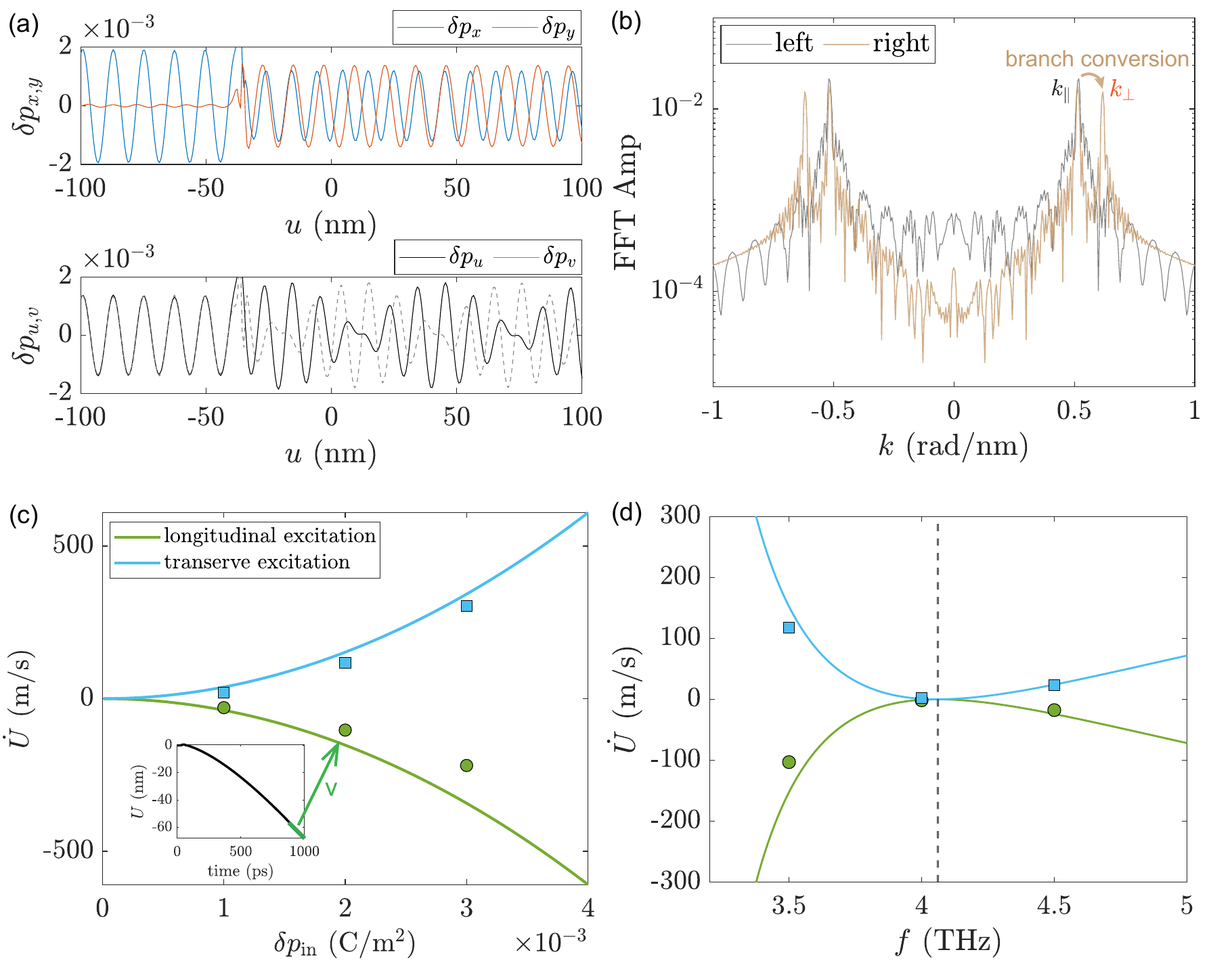}
 \caption{
(a) Spatial profiles of $\delta p_{x,y}$ and $\delta p_{u,v}$ at $t=500$ ps after activating a longitudinal ferron source with $f=3.5$ THz at $u_s=-200$ nm.  
(b) Fourier spectrum of $\delta p_u(u,t)$ on the left (gray) and right (brown) sides of the DW. The extra peak on the right indicates partial conversion of the incoming longitudinal mode to a transverse mode with different wave number.
(c) DW velocities as a function of amplitude for longitudinal (green) and transverse (blue) incident modes.
    Symbols are obtained from numerical simulations of the DW velocity. Inset: time dependent DW position under the conditions in (a).
(d) DW velocity as a function of incident ferron frequencies for $\delta p_{\rm in}=0.002$ C/m$^2$.}\label{F3}
\end{figure}

An anisotropy $\kappa=2.9$ slightly splits the longitudinal and transverse waves  as shown in Fig.~\ref{F2}(b) that at a fixed frequency propagate with different wave vectors, $k_{\parallel}$ and $k_{\perp}$. The snapshot of the time-dependent amplitudes in Fig.~\ref{F3}(a) shows that the DW is still nearly transparent for the incoming longitudinal mode. The mode conversion in the rotated coordinate system combined with the phase shift leads to the beating pattern in $\delta p_{u,v}$. Figure \ref{F3}(b) is the spatiotemporal Fourier transform of the simulated time series on the left and right side of the DW. The additional  peak in the spectral function of the transmitted amplitude is direct evidence of the mode splitting. The incoming and outgoing currents carry now different momentum fluxes, and their difference causes a force $\Delta J_P$, i.e., the difference between the incoming and outgoing ferron momentum fluxes, on the DW. The mode conversion in non-collinear textures therefore leads to a finite forces already in linear response to a ferron current, in contrast to the Ising-wall case \cite{Yang2026}.

To estimate this force, we adopt the lowest-order collective-coordinate ansatz $p_v(u,t)=p_v^0[u-U(t)]$, where $U(t)$ denotes the DW center. Projecting the LKT equation onto the translational zero mode, i.e.,  multiplying it by $\partial_u p_v^0$ and integrating over $u$, yields
\begin{equation}
M_{\rm DW}\ddot U+\Gamma_{\rm DW}\dot U=\Delta J_P ,
\end{equation}
where $M_{\rm DW}=m_p\int du (\partial_u p_v^0)^2=2m_p p_s^2/(3\Delta_{\rm I})$ and $\Gamma_{\rm DW}=\gamma\int du (\partial_u p_v^0)^2=2\gamma p_s^2/(3\Delta_{\rm I})$. Disregarding higher harmonics, the momentum flux of $\delta p_\mu(u,t)=A_\mu\cos(k_\mu u-\omega t)$ reads $J_{P,\mu}=\frac12 g k_\mu^2 A_\mu^2$, where the subscript \(\mu\) labels the incoming, reflected, and  outgoing modes. At the saturation velocity \(\dot U_\mathrm{sat}\) in the steady state the driving force equals the braking by the viscous DW damping, giving 
\begin{equation}
\dot U_\mathrm{sat}=
\frac{3g\Delta_{\rm I}}{4\gamma p_s^2}
\Big(k_{\rm in}^2A_{\rm in}^2-\sum_{\mu\in{\rm ref,out}}k_\mu^2A_\mu^2\Big),
\end{equation}
where the sum includes all reflected and transmitted modes. The outgoing amplitudes $A_{\parallel}$ and $A_{\perp}$ depend on the beam-splitting ratio $\chi_{\parallel}$ and fulfill the energy conservation relation $A_{\mathrm{in}}^2=A_{\parallel}^2+A_{\perp}^2+A_{\rm ref}^2$ under weak damping.

At $\kappa=2.9$, the scattering is still nearly reflectionless, which allows us to disregard the reflected contribution ($A_{\rm ref}=0$). When using the analytical beam-splitting ratio obtained at $\kappa=3$ for the transmitted amplitudes the velocity reduces to
\begin{equation}\label{DWv1}
\dot U_\mathrm{sat} \approx\frac{3g\Delta_{\rm I}}{4\gamma p_s^2}
(k_{\mathrm{in}}^2 A_{\mathrm{in}}^2-k_{\parallel}^2 A_{\parallel}^2-k_{\perp}^2 A_{\perp}^2),
\end{equation}
which is proportional to \(A_{\mathrm{in}}^2\), the incident wave intensity. Since $k_\perp>k_\parallel$ at the same frequency, longitudinal-to-transverse conversion ($k_{\mathrm{in}}=k_\parallel$) increases the outgoing momentum flux, therefore $\dot U_{\rm sat}<0$, whereas transverse-to-longitudinal conversion ($k_{\mathrm{in}}=k_\perp$) decreases it and $\dot U_{\rm sat}>0$.

We can test the above estimates by numerically simulating the DW model in Fig.~\ref{F2}(a) for $N=2001$ dipoles when driven by locally oscillating fields $E_x=E_{x0}\delta(u-u_s)\sin(2\pi f t)$, $E_y=0$ as well as $E_y=E_{y0}\delta(u-u_s)\sin(2\pi f t)$, $E_x=0$ with $f=3.5$ THz at $u_s=-200$ nm. Figure~\ref{F3}(c) shows the saturation velocity in both cases. The estimates from Eq.~\eqref{DWv1} agree well with the numerical results (dots). A transverse source reverses the sign of the momentum-flux imbalance of a longitudinal one and the DW moves in opposite directions, as shown by the blue vs. green curves and dots in Fig.~\ref{F3}(c). The non-monotonic dependence of the DW velocity on the driving frequency in Fig.~\ref{F3}(d) reflects the dependence of the phase shift on the wave number in Fig.~\ref{F2}(c).

This mechanism is reminiscent of the mechanical action of polarized light on a birefringent medium, i.e., changing the polarization state of light transfers angular momentum to the material as a \textit{torque} \cite{Beth1936}. In contrast, the transformation between the phonon branches and the resulting imbalance of linear momentum flux is transferred to the DW as a \textit{force}. 

\begin{figure}
  \centering
  \includegraphics[width=0.49\textwidth]{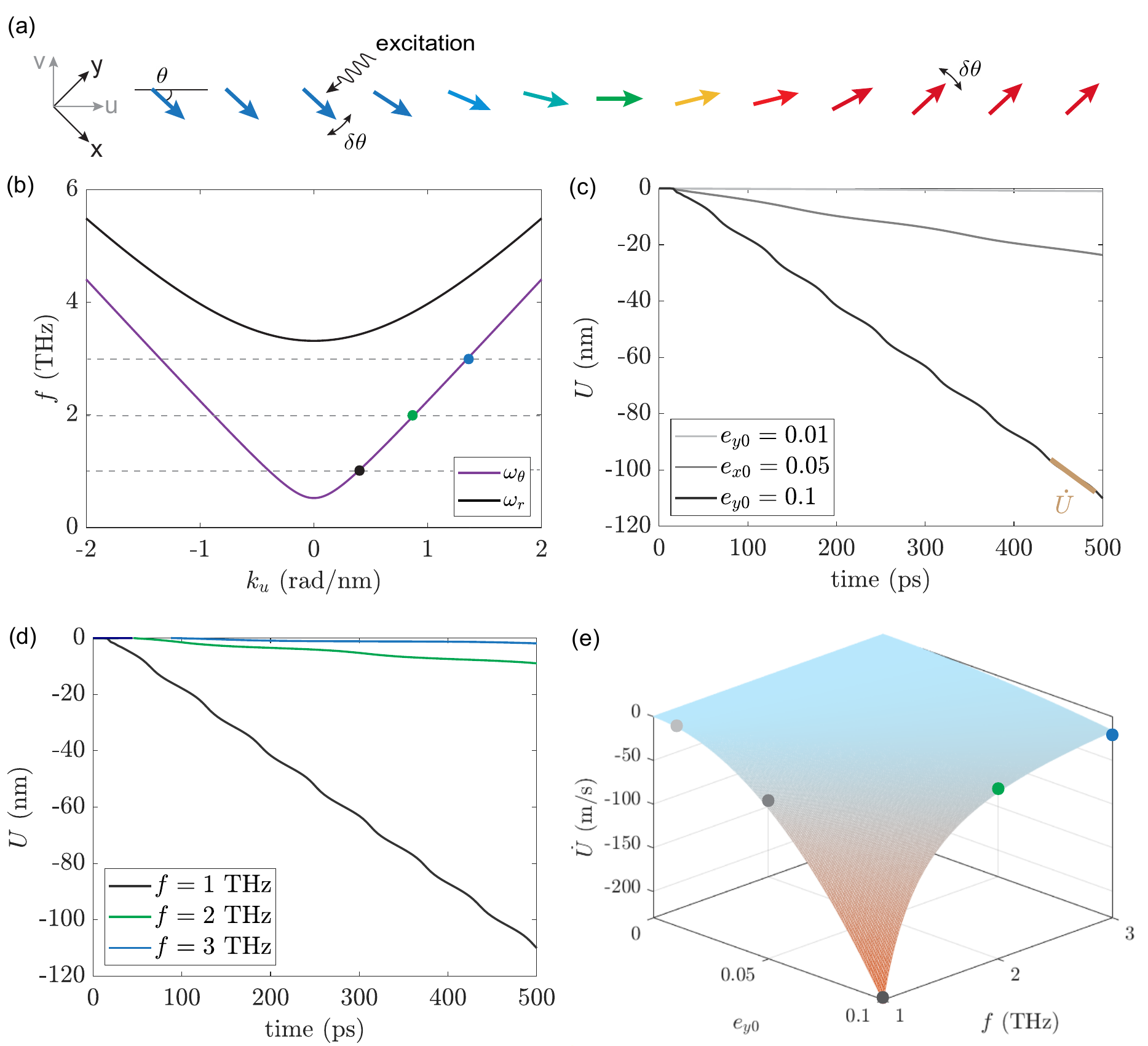}
\caption{(a) Schematic of a transverse phonon ($\theta$-mode).
(b) Frequency dispersions of the longitudinal ferrons (black) and $\theta$-mode (purple)  in the limit of weak anisotropy ($\kappa=1.05$).
(c) DW position versus time for different drive amplitudes $e_{y0}=E_{y0}/(|\alpha|p_s)$ at $f=1$ THz. The gray segment marks the linear-fitting window used to approximate the steady state DW velocity at a source-wall distance of $L=100$ nm.
(d) DW position versus time for different driving frequencies at fixed amplitude $e_{y0}=0.1$.
(e) DW velocity dependence on the drive frequency and electric-field amplitude. Dots correspond to the DW velocities extracted from (c) and (d) in the vicinity of $u=-100$ nm.
}\label{F4}
\end{figure}

{\color{blue} {\it II. Fixed-modulus wall.}}
The analysis breaks down for an isotropic ferroelectric with $\kappa=1$ since the DW width diverges. Here we focus on a weak-anisotropy scenario with $\kappa=1.05$. The polarizations then rotates with nearly constant modulus, as shown in Fig.~\ref{F4}(a). The longitudinal ferrons form a high-frequency branch [Fig.~\ref{F4}(b)], while a sine-Gordon equation governs the dynamics of the angular degree of freedom $\theta$. With a harmonic drive at position $u_s$ \cite{SM} at frequency in the lower branch, the dynamical equation reads
\begin{equation}
\theta_{tt}+\tilde\gamma\theta_t-c^2\theta_{uu}
-\frac14\Omega_0^2\sin4\theta
=h\cos(\omega t)\delta(u-u_s),
\end{equation}
where $\tilde\gamma=\gamma/m_p$, $c^2=g/m_p$, $\Omega_0^2=p_s^2(\beta_2-\beta_1)/m_p$, and $h=aE_{y0}/(m_p p_s)$ with lattice constant $a$.
 
When linearizing the problem for small fluctuations $\delta\theta(u,t)$  and disregard damping, we arrive at a reflectionless P\"oschl-Teller scattering problem equivalent to that of a sine-Gordon kink, which means that the force on the DW vanishes. However, the situation changes when damping is included \cite{Kivshar1993}.

Separating the field into a wall coordinate $\Theta (u)$ that rotates slowly in the wide DW and a quickly oscillating component $\eta$, the time-averaged wave intensity $\langle \eta^2(u)\rangle_t$ affects the slow wall dynamics. The resulting equation of motion for the collective-coordinate is
\begin{equation}
M_\theta\ddot U+\tilde\gamma M_\theta\dot U=2c\Omega_0\frac{d}{dU}\langle\eta^2(U)\rangle_t,
\end{equation}
where $M_\theta=\Omega_0/(2c)$. This expression remains valid in the presence of the depolarization field.

In a dissipative uniform medium, a monochromatic local source at $u_s$ generates the response
\begin{equation}
\phi(u,t)=\mathrm{Re}\left[\frac{h}{2c^2q}e^{-q|u-u_s|}e^{-i\omega t}\right],
\end{equation}
where $q=c^{-1}\sqrt{\Omega_0^2-\omega^2-i\tilde\gamma\omega}$ is a complex wave vector \cite{SM}. In the calculations, we set the dimensionless viscous damping
$\gamma/\sqrt{m_p|\alpha|}=10^{-4}$, corresponding to $\gamma=2\times10^{-9}\ {\rm J\,m\,s/C^2}$ and a ferron mean free path $\ell\simeq11.2\,\mu{\rm m}$ at $f=1\,{\rm THz}$. The time-averaged intensity 
$\langle\eta^2(u)\rangle_t=\frac12|\phi(u,t)|^2$,
leads to the DW velocity
\begin{equation}\label{DWv2}
\dot U=-\frac{g h^2}{\gamma c^4}\frac{\mathrm{Re}(q)}{|q|^2}\operatorname{sgn}(U-u_s)\exp[-2\mathrm{Re}(q)|U-u_s|].
\end{equation}
The intensity gradient of the oscillatory ferron background therefore pulls the wall toward the source.

This source-seeking motion is quantified in Figs.~\ref{F4}(c) and \ref{F4}(d), where we show the computed results for a source at $u_s=-200$ nm with $E_y=E_{y0}\delta(u-u_s)\sin(2\pi f t)$ and $E_x=0$. The DW velocity is extracted from a linear fit to $U(t)$ near $U=-100$ nm, corresponding to a source-wall distance $L=100$ nm, with fitting window indicated by the brown segment in Fig.~\ref{F4}(c) . For other parameters, the simulations are run longer so that the fit is carried out at the same wall position. The drift speed increases with the drive amplitude and quickly decreases with frequency, which also renders a drive at the upper mode frequency very inefficient. Figure~\ref{F4}(e) demonstrates agreement between Eq.~\eqref{DWv2} and the numerical results and identifies the driving mechanism to be a dissipation-induced gradient force rather than linear-momentum transfer. 

{\color{blue} {\it Discussion and conclusions.}}
The effective model used here is derived from a sixth-order Landau free energy for BaTiO$_3$ \cite{SM}, in which the Landau coefficients are stress-free since obtained from minimizing the full electromechanical free energy with respect to the homogeneous strain. Elastic and electrostrictive effects are therefore already built into the free energy. The quartic model used in the main text is an effective expansion near the ferroelectric minimum and thus retains these underlying electromechanical contributions. 

The dipole-dipole interaction is not  included in our model above, but we analyze its effects in the Supplemental Material \cite{SM}. For waves propagating normal to the wall, the dynamic depolarization field couples to the polarization component parallel to the propagation direction $\hat u$. Since $\delta p_u$ is a linear combination of the local longitudinal and transverse modes, the depolarization field causes both a splitting and mixing of the local modes. When the splitting is small, the phase-induced branch-conversion mechanism discussed in the main text remains active. This regime can be accessed in thin films and an environment with high dielectric constant. Otherwise, the hybridization and force is weakened. In this limit, a transverse optical phonon can still drive the wall through the dissipation-induced gradient force associated with the nonuniform ferron background.

Mode conversion at interfaces underlies longitudinal-transverse elastic-wave transduction and spin-wave refraction at magnetic DWs \cite{Dong2022,Hioki2020Snell}. In contrast to these effectively fixed interfaces, the ferroelectric DW plays a dual role as both a mode converter and a mobile collective coordinate, converting the branch-dependent momentum-flux imbalance into wall motion and suggesting a general route to wave-controlled mobile interfaces.

The predicted source-seeking motion suggests that DWs can be spatially dragged by moving a local ferron source, such as an optical spot or line. A similar fixed-modulus picture is also suitable for a $180^\circ$ described by a twofold anisotropy in bilayer h-BN \cite{He2024ActaMat}. The predicted dynamics should be accessible experimentally using local ac excitation of coherent ferrons \cite{Choe2026,Zhang2025,Subedi2026} together with time-resolved probes of DW motion \cite{Gruverman2019}. Related behavior may arise in other multicomponent polar textures, including polar vortices \cite{Yadav2016,SanchezSantolino2024}, antivortices \cite{Abid2021}, and skyrmion bubbles \cite{Xue2025}. 

In summary, current-driven ferroelectric DW dynamics exhibits richer behavior than its magnetic counterpart, as summarized in Supplemental Material \cite{SM}. We predict that in general $90^\circ$ DWs mix the longitudinal ferron and the transverse optical phonon modes. Due to the mode conversion, incoming polarization and momentum currents are not conserved. The resulting forces may drive the dynamics of the ferroelectric texture such as DW motion. The forces vanish for the balanced DW texture with in-plane polarizations \(p_x+p_y=p_s\) that forms the equilibrium texture for a special cubic anisotropy and for polarizations that rotated with constant modulus in weakly anisotropic ferroelectrics. In these limits we predict that dissipation-induced gradient forces take over. Our results identify the key ingredients that control wave-driven DW motion in multicomponent ferroelectrics. DWs may act as beam splitters in ferronic logic circuits or as interferometric electric field detectors that operate at THz frequencies. 

\begin{acknowledgments}
This work is supported by JSPS KAKENHI Grants No. 22H04965 and No. 24H02231.
\end{acknowledgments}


\begin{thebibliography}{99}
\bibitem{Catalan2012}
G.~Catalan, J.~Seidel, R.~Ramesh, and J.~F.~Scott,
Domain wall nanoelectronics,
\href{https://doi.org/10.1103/RevModPhys.84.119}{Rev.\ Mod.\ Phys.\ \textbf{84}, 119 (2012)}.

\bibitem{CherifiHertel2017}
S.~Cherifi-Hertel, H.~Bulou, R.~Hertel, G.~Taupier, K.~D.~H.~Dorkenoo,
C.~Andreas, J.~Guyonnet, I.~Gaponenko, K.~Gallo, and P.~Paruch,
Non-Ising and chiral ferroelectric domain walls revealed by nonlinear optical microscopy,
\href{https://doi.org/10.1038/ncomms15768}{Nat.\ Commun.\ \textbf{8}, 15768 (2017)}.

\bibitem{Mantri2021}
S.~Mantri and J.~Daniels,
Domain walls in ferroelectrics,
\href{https://doi.org/10.1111/jace.17555}{J.\ Am.\ Ceram.\ Soc.\ \textbf{104}, 1619 (2021)}.

\bibitem{Nataf2020}
G.~F.~Nataf, M.~Guennou, J.~M.~Gregg, D.~Meier, J.~Hlinka, E.~K.~H.~Salje, and J.~Kreisel,
Domain-wall engineering and topological defects in ferroelectric and ferroelastic materials,
\href{https://doi.org/10.1038/s42254-020-0235-z}{Nat.\ Rev.\ Phys.\ \textbf{2}, 634 (2020)}.

\bibitem{He2024ActaMat}
R.~He, B.~Zhang, H.~Wang, L.~Li, P.~Tang, G.~Bauer, and Z.~Zhong,
Ultrafast switching dynamics of the ferroelectric order in stacking-engineered ferroelectrics,
\href{https://doi.org/10.1016/j.actamat.2023.119416}{Acta Mater.\ \textbf{262}, 119416 (2024)}.

\bibitem{Wei2016}X.-K.~Wei, C.-L.~Jia, T.~Sluka, B.-X.~Wang, Z.-G.~Ye, and N.~Setter, N\'eel-like domain walls in ferroelectric Pb(Zr,Ti)O$_3$ single crystals,
\href{https://doi.org/10.1038/ncomms12385}{Nat.\ Commun.\ \textbf{7}, 12385 (2016)}.


\bibitem{Cao1991} W.~Cao and L.~E.~Cross,
Theory of tetragonal twin structures in ferroelectric perovskites with a first-order phase transition,
\href{https://doi.org/10.1103/PhysRevB.44.5}{Phys.\ Rev.\ B \textbf{44}, 5 (1991)}.

\bibitem{Hlinka2006} J.~Hlinka and P.~M\'arton, Phenomenological model of a 90$^\circ$ domain wall in BaTiO$_3$-type ferroelectrics,
\href{https://doi.org/10.1103/PhysRevB.74.104104}{Phys.\ Rev.\ B \textbf{74}, 104104 (2006)}.

\bibitem{Morikawa2023}
D.~Morikawa, Y.~Noguchi, and K.~Tsuda,
Direct observation of rotation of polarization at 90-degree domain walls in BaTiO$_3$,
\href{https://doi.org/10.35848/1347-4065/acde5b}{Jpn.\ J.\ Appl.\ Phys.\ \textbf{62}, SM1003 (2023)}.

\bibitem{Karthik2012}
J.~Karthik, J.~C.~Agar, A.~R.~Damodaran, and L.~W.~Martin,
Effect of 90$^\circ$ Domain Walls and Thermal Expansion Mismatch on the Pyroelectric Properties of Epitaxial PbZr$_{0.2}$Ti$_{0.8}$O$_3$ Thin Films,
\href{https://doi.org/10.1103/PhysRevLett.109.257602}{Phys.\ Rev.\ Lett.\ \textbf{109}, 257602 (2012)}.

\bibitem{Matzen2014}
S.~Matzen, O.~Nesterov, G.~Rispens, J.~A.~Heuver, M.~Biegalski, H.~M.~Christen, and B.~Noheda,
Super switching and control of in-plane ferroelectric nanodomains in strained thin films,
\href{https://doi.org/10.1038/ncomms5415}{Nat.\ Commun.\ \textbf{5}, 4415 (2014)}.


\bibitem{Tang2014}
Y.~L.~Tang, Y.~L.~Zhu, Y.~J.~Wang, W.~Y.~Wang, Y.~B.~Xu, W.~J.~Ren, Z.~D.~Zhang, and X.~L.~Ma,
Atomic-scale mapping of dipole frustration at 90$^\circ$ charged domain walls in ferroelectric PbTiO$_3$ films,
\href{https://doi.org/10.1038/srep04115}{Sci.\ Rep.\ \textbf{4}, 4115 (2014)}.

\bibitem{Hirohashi2003}
J.~Hirohashi, K.~Yamada, H.~Kamio, and S.~Shichijyo,
Fabrication of 90$^\circ$ Domain Structures in KNbO$_3$ Single Crystals,
\href{https://doi.org/10.1080/00150190303806}{Ferroelectrics \textbf{281}, 29 (2003)}.



\bibitem{Zheng2023} Y.~Zheng, Y.~Zhang, T.~Xin, Y.~Xu, S.~Qu, J.~Zheng, Z.~Gao, Q.~Zhong, Y.~Wang, X.~Feng, \textit{et al.},
Direct atomic-scale visualization of the 90$^\circ$ domain walls and their migrations in Hf$_{0.5}$Zr$_{0.5}$O$_2$ ferroelectric thin films,
\href{https://doi.org/10.1016/j.mtnano.2023.100406}{Materials Today Nano \textbf{24}, 100406 (2023)}.

\bibitem{Wen2026}
S.~Wen, R.-C.~Peng, X.~Cheng, M.~Liao, and Y.~Zhou,
The dual role of 90$^\circ$ domain walls in ferroelectric switching of Hf$_{0.5}$Zr$_{0.5}$O$_2$ thin films: Insights from phase-field simulations,
\href{https://doi.org/10.1038/s41524-026-02028-7}{npj Comput.\ Mater. \textbf{12}, 158 (2026)}.

\bibitem{Fei2016}
R.~Fei, W.~Kang, and L.~Yang,
Ferroelectricity and Phase Transitions in Monolayer Group-IV Monochalcogenides,
\href{https://doi.org/10.1103/PhysRevLett.117.097601}{Phys.\ Rev.\ Lett.\ \textbf{117}, 097601 (2016)}.

\bibitem{BarrazaLopez2021}
S.~Barraza-Lopez, B.~M.~Fregoso, J.~W.~Villanova, S.~S.~P.~Parkin, and K.~Chang,
Colloquium: Physical properties of group-IV monochalcogenide monolayers,
\href{https://doi.org/10.1103/RevModPhys.93.011001}{Rev.\ Mod.\ Phys.\ \textbf{93}, 011001 (2021)}.

\bibitem{Yan2022} Y.~Yan, M.~Xiang, X.~Wang, T.~Xu, and F.~Xuan,
Ferroelectric domain wall in two-dimensional GeS,
\href{https://doi.org/10.1063/5.0094689}{J.\ Appl.\ Phys.\ \textbf{132}, 074302 (2022)}.

\bibitem{Gao2013} P.~Gao, J.~Britson, J.~R.~Jokisaari, C.~T.~Nelson, S.-H.~Baek, Y.~Wang, C.-B.~Eom, L.-Q.~Chen, and X.~Pan, Atomic-scale mechanisms of ferroelastic domain-wall-mediated ferroelectric switching, \href{https://doi.org/10.1038/ncomms3791}{Nat.\ Commun.\ \textbf{4}, 2791 (2013)}.

\bibitem{Shin2007} Y.-H.~Shin, I.~Grinberg, I.-W.~Chen, and A.~M.~Rappe, Nucleation and growth mechanism of ferroelectric domain-wall motion,
\href{https://doi.org/10.1038/nature06165}{Nature \textbf{449}, 881 (2007)}.

\bibitem{Kagawa2016}
F.~Kagawa, N.~Minami, S.~Horiuchi, and Y.~Tokura,
Athermal domain-wall creep near a ferroelectric quantum critical point,
\href{https://doi.org/10.1038/ncomms10675}{Nat.\ Commun.\ \textbf{7}, 10675 (2016)}.

\bibitem{Jo2009}
J.~Y.~Jo, S.~M.~Yang, T.~H.~Kim, H.~N.~Lee, J.-G.~Yoon, S.~Park, Y.~Jo, M.~H.~Jung, and T.~W.~Noh,
Nonlinear Dynamics of Domain-Wall Propagation in Epitaxial Ferroelectric Thin Films,
\href{https://doi.org/10.1103/PhysRevLett.102.045701}{Phys.\ Rev.\ Lett.\ \textbf{102}, 045701 (2009)}.

\bibitem{Sharma2019}
P.~Sharma, P.~Schoenherr, and J.~Seidel,
Functional Ferroic Domain Walls for Nanoelectronics,
\href{https://doi.org/10.3390/ma12182927}{Materials \textbf{12}, 2927 (2019)}.

\bibitem{Meier2022}
D.~Meier and S.~M.~Selbach,
Ferroelectric domain walls for nanotechnology,
\href{https://doi.org/10.1038/s41578-021-00375-z}{Nat.\ Rev.\ Mater.\ \textbf{7}, 157 (2022)}.

\bibitem{Junquera2023}
J.~Junquera, Y.~Nahas, S.~Prokhorenko, L.~Bellaiche, J.~\'Iniguez, D.~G.~Schlom, L.-Q.~Chen, S.~Salahuddin, and D.~A.~Muller, {\it et al.},
Topological phases in polar oxide nanostructures,
\href{https://doi.org/10.1103/RevModPhys.95.025001}{Rev.\ Mod.\ Phys.\ \textbf{95}, 025001 (2023)}.

\bibitem{McCluskey2025}
C.~J.~McCluskey, K.~M.~Holsgrove, and J.~M.~Gregg,
Perspective: Domain wall nanoelectronics,
\href{https://doi.org/10.1063/5.0279244}{J.\ Appl.\ Phys.\ \textbf{138}, 050901 (2025)}.


\bibitem{Yamaguchi2004}
A.~Yamaguchi, T.~Ono, S.~Nasu, K.~Miyake, K.~Mibu, and T.~Shinjo,
Real-Space Observation of Current-Driven Domain Wall Motion in Submicron Magnetic Wires,
\href{https://doi.org/10.1103/PhysRevLett.92.077205}{Phys.\ Rev.\ Lett.\ \textbf{92}, 077205 (2004)}.

\bibitem{Parkin2008}
S.~S.~P.~Parkin, M.~Hayashi, and L.~Thomas,
Magnetic Domain-Wall Racetrack Memory,
\href{https://doi.org/10.1126/science.1145799}{Science \textbf{320}, 190 (2008)}.

\bibitem{Yan2011}P. Yan, X. Wang, and X. R. Wang, All-Magnonic Spin-Transfer Torque and Domain Wall Propagation, 
\href{https://doi.org/10.1103/PhysRevLett.107.177207}{Phys. Rev. Lett. \textbf{107}, 177207 (2011)}.   

\bibitem{Hinzke2011}
D.~Hinzke and U.~Nowak,
Domain Wall Motion by the Magnonic Spin Seebeck Effect,
\href{https://doi.org/10.1103/PhysRevLett.107.027205}{Phys.\ Rev.\ Lett.\ \textbf{107}, 027205 (2011)}.
 
\bibitem{Wang2012Emission}
X.~S.~Wang, P.~Yan, Y.~H.~Shen, G.~E.~W.~Bauer, and X.~R.~Wang,
Domain Wall Propagation through Spin Wave Emission,
\href{https://doi.org/10.1103/PhysRevLett.109.167209}{Phys.\ Rev.\ Lett.\ \textbf{109}, 167209 (2012)}.

\bibitem{Kim2012}
J.-S. Kim, M. Stärk, M. Kläui, J. Yoon, C.-Y. You, L. Lopez-Diaz, and E. Martinez,
Interaction between propagating spin waves and domain walls on a magnetic nanowire,
\href{https://doi.org/10.1103/PhysRevB.85.174428}{Phys.\ Rev.\ B \textbf{85}, 174428 (2012)}.

\bibitem{Schlickeiser2014}
F.~Schlickeiser, U.~Ritzmann, D.~Hinzke, and U.~Nowak,
Role of Entropy in Domain Wall Motion in Thermal Gradients,
\href{https://doi.org/10.1103/PhysRevLett.113.097201}{Phys.\ Rev.\ Lett.\ \textbf{113}, 097201 (2014)}.
  
\bibitem{Wang2015DMI}
W.~Wang, M.~Albert, M.~Beg, M.-A.~Bisotti, D.~Chernyshenko, D.~Cort\'es-Ortu\~no, I.~Hawke, and H.~Fangohr,
Magnon-Driven Domain-Wall Motion with the Dzyaloshinskii-Moriya Interaction,
\href{https://doi.org/10.1103/PhysRevLett.114.087203}{Phys.\ Rev.\ Lett.\ \textbf{114}, 087203 (2015)}.

\bibitem{Kim2015}
S.~K.~Kim and Y.~Tserkovnyak,
Landau-Lifshitz theory of thermomagnonic torque,
\href{https://doi.org/10.1103/PhysRevB.92.020410}
{Phys. Rev. B \textbf{92}, 020410 (2015)}.

\bibitem{Kim2019Unidirectional}
K.-W.~Kim, S.-W.~Lee, J.-H.~Moon, G.~Go, A.~Manchon,
H.-W.~Lee, K.~Everschor-Sitte, and K.-J.~Lee,
Unidirectional Magnon-Driven Domain Wall Motion Due to the Interfacial Dzyaloshinskii-Moriya Interaction,
\href{https://doi.org/10.1103/PhysRevLett.122.147202}{Phys.\ Rev.\ Lett.\ \textbf{122}, 147202 (2019)}.

\bibitem{Lan2022}
J.~Lan and J.~Xiao,
Spin wave driven domain wall motion in easy-plane ferromagnets: A particle perspective,
\href{https://doi.org/10.1103/PhysRevB.106.L020404}{Phys.\ Rev.\ B \textbf{106}, L020404 (2022)}.

\bibitem{Bauer2021}
G.~E.~W.~Bauer, R.~Iguchi, and K.-i.~Uchida, Theory of Transport in Ferroelectric Capacitors, 
\href{https://doi.org/10.1103/PhysRevLett.126.187603}{Phys. Rev. Lett. \textbf{126}, 187603 (2021)}.

\bibitem{Tang2022}
P.~Tang, R.~Iguchi, K.-i.~Uchida, and G.~E.~W.~Bauer, Excitations of the ferroelectric order,
\href{https://doi.org/10.1103/PhysRevB.106.L081105}{Phys.\ Rev.\ B \textbf{106}, L081105 (2022)}.

\bibitem{Tang2024Ferron}P.~Tang and G.~E.~W.~Bauer, Electric analog of magnons in order-disorder ferroelectrics,
\href{https://doi.org/10.1103/PhysRevB.109.L060301}{Phys.\ Rev.\ B \textbf{109}, L060301 (2024)}.

\bibitem{Adachi2023}
H.~Adachi, N.~Ikeda, and E.~Saitoh,
Ginzburg-Landau action and polarization current in an excitonic insulator model of electronic ferroelectricity,
\href{https://doi.org/10.1103/PhysRevB.107.155142}{Phys.\ Rev.\ B \textbf{107}, 155142 (2023)}.

\bibitem{RodriguezSuarez2024}
R.~L.~Rodr\'iguez-Su\'arez, X.-H.~Zhou, C.~Y.~Cai, P.~Tang, T.~Yu, G.~E.~W.~Bauer, and S.~M.~Rezende,
Surface and volume modes of polarization waves in ferroelectric films,
\href{https://doi.org/10.1103/PhysRevB.109.134307}{Phys.\ Rev.\ B \textbf{109}, 134307 (2024)}.

\bibitem{Shen2025}K. Shen, P. Tang, X. Chen, Y. Gao, Y. Fan, Z. Guo, Y. Wei, H. Jiang, X. Zhang, M. Wang, \textit{et al.}, Observation of ferron transport in ferroelectrics, \href{https://doi.org/10.48550/arXiv.2505.24419}{arXiv:2505.24419 (2025)}.

\bibitem{Choe2026}J. Choe, T. Handa, C. Huang, A. Koch Liston, J. Cox, J. Stensberg, Y. Hong, D. G. Chica, D. Xu, F. Tay, \textit{et al.}, Observation of coherent ferron emission and propagation, \href{https://doi.org/10.1038/s41563-026-02597-4}{Nat. Mater. (2026)}.

\bibitem{Zhang2025} B. Zhang, R. Duan, S. S. Mishra, S. Jana, J. Kim, T. T. Caiwei, Y. J. Tan, W. Wang, P. T. C. Ietro, Z. Liu, and R. Singh, Electric-field control of coherent ferron oscillations,   \href{https://doi.org/10.1038/s41567-026-03406-0}{Nat. Phys. (2026)}.

\bibitem{Subedi2026}S.~Subedi, W.~Fang, F.~Fei, Z.~Zhai, J.~P.~Rollins, C.~Fox, A.~Drew, B.~Lv, Y.~Ping, and J.~Xiao, Electrically switchable ferron upconversion in a van der Waals ferroelectric, \href{https://doi.org/10.48550/arXiv.2603.19394}{arXiv:2603.19394 (2026)}.

\bibitem{Manna2026}M.~Manna, F.~Yang, S.~Chaudhuri, S.~Shenoy, G.-D.~Zhao, M.~Tanwani, S.~P.~Reddy, P.~J.~Hays, S.~Sankar, C.~de~Souza~Coutinho~Nogueira, et al.,
Probing (sub)nanoscale ferrons in an electron microscope, \href{https://arxiv.org/abs/2609.13625}{arXiv:2609.13625 (2026)}.

\bibitem{Yang2026} H.~Yang, P.~Yan, and G.~E.~W.~Bauer, Polarization transfer force on ferroelectric domain walls,
\href{https://doi.org/10.1103/hkmj-6skv}{Phys. Rev. B \textbf{113}, L220103 (2026)}.

\bibitem{Tagantsev2009}A. K. Tagantsev, L. Eric Cross, and J. Fousek, {\it Domains in Ferroic crystals and thin films}, (Springer, New York, 2010).

\bibitem{Ishibashi2002}Y.~Ishibashi and E.~Salje, A Theory of Ferroelectric 90 Degree Domain Wall,
\href{https://doi.org/10.1143/JPSJ.71.2800}{J. Phys. Soc. Jpn. \textbf{71}, 2800 (2002)}.

\bibitem{Azuma2023} H.~Azuma, S.~Ogata, R.~Kobayashi, M.~Uranagase, T.~Tsuzuki, D.~Durdiev, and F.~Wendler, Microscopic structure and migration of 90$^\circ$ ferroelectric domain wall in BaTiO$_3$ determined via molecular dynamics simulations, \href{https://doi.org/10.1063/5.0138489}{J.\ Appl.\ Phys.\ \textbf{133}, 104101 (2023)}.


\bibitem{Tani1969} K.~Tani, Dynamics of Displacive-Type Ferroelectrics,
\href{https://doi.org/10.1143/JPSJ.26.93}{J.\ Phys.\ Soc.\ Jpn.\ \textbf{26}, 93 (1969)}.

\bibitem{Sivasubramanian2004}S.~Sivasubramanian, A.~Widom, and Y.~N.~Srivastava,
Physical kinetics of ferroelectric hysteresis,
\href{https://doi.org/10.1080/00150190490442173}{Ferroelectrics \textbf{300}, 43 (2004)}.

\bibitem{SM}See Supplemental Material at \url{http://link.aps.org/supplemental/xxx} for material parameters, analytical derivations, analysis of the dynamic depolarization field, experimental considerations, and a comparison of ferron-driven motion for different domain-wall configurations.
    
    
\bibitem{Beth1936} R.~A.~Beth, Mechanical Detection and Measurement of the Angular Momentum of Light,
\href{https://doi.org/10.1103/PhysRev.50.115}{Phys.\ Rev.\ \textbf{50}, 115 (1936)}.


\bibitem{Kivshar1993}Y.~S.~Kivshar, O.~H.~Olsen, M.~R.~Samuelsen, and A.~V.~Ustinov, Fluxon interaction with external rf radiation in Josephson junctions,
\href{https://doi.org/10.1103/PhysRevB.47.5212}{Phys.\ Rev.\ B \textbf{47}, 5212 (1993)}.



\bibitem{Dong2022} H.-W.~Dong, S.-D.~Zhao, M.~Oudich, C.~Shen, C.~Zhang, L.~Cheng, Y.-S.~Wang, and D.~Fang, Reflective metasurfaces with multiple elastic mode conversions for broadband underwater sound absorption, \href{https://doi.org/10.1103/PhysRevApplied.17.044013} {Phys. Rev. Applied \textbf{17}, 044013 (2022)}.

\bibitem{Hioki2020Snell} T.~Hioki, R.~Tsuboi, T.~H.~Johansen, Y.~Hashimoto, and E.~Saitoh, Snell's law for spin waves at a $90^\circ$ magnetic domain wall, \href{https://doi.org/10.1063/1.5141864} {Appl. Phys. Lett. \textbf{116}, 112402 (2020)}. 

\bibitem{Gruverman2019}
A.~Gruverman, M.~Alexe, and D.~Meier, Piezoresponse force microscopy and nanoferroic phenomena, \href{https://doi.org/10.1038/s41467-019-09650-8}{Nat.\ Commun.\ \textbf{10}, 1661 (2019)}.

\bibitem{Yadav2016} A.~K.~Yadav, C.~T.~Nelson, S.~L.~Hsu, Z.~Hong, J.~D.~Clarkson, C.~M.~Schlep\"utz, A.~R.~Damodaran, P.~Shafer, E.~Arenholz, and L.~R.~Dedon, \textit{et al.},
Observation of polar vortices in oxide superlattices, \href{https://doi.org/10.1038/nature16463}{Nature \textbf{530}, 198 (2016)}.

\bibitem{SanchezSantolino2024} G.~S\'anchez-Santolino, V.~Rouco, S.~Puebla, H.~Aramberri, V.~Zamora, M.~Cabero, F.~A.~Cuellar, C.~Munuera, F.~Mompean, and M.~Garc{\'\i}a-Hern{\'a}ndez, \textit{et al.}, A 2D ferroelectric vortex pattern in twisted BaTiO$_3$ freestanding layers, \href{https://doi.org/10.1038/s41586-023-06978-6}{Nature \textbf{626}, 529 (2024)}.

\bibitem{Abid2021}
A.~Y.~Abid, Y.~Sun, X.~Hou, C.~Tan, X.~Zhong, R.~Zhu, H.~Chen, K.~Qu, Y.~Li, and M.~Wu, \textit{et al.},
Creating polar antivortex in PbTiO$_3$/SrTiO$_3$ superlattice,
\href{https://doi.org/10.1038/s41467-021-22356-0}{Nat.\ Commun.\ \textbf{12}, 2054 (2021)}.


\bibitem{Xue2025}
F.~Xue, C.~Zhang, S.~Zheng, P.~Tong, B.~Wang, Y.~Peng, Z.~Wang, H.~Xu, Y.~He, and H.~Zhou, \textit{et al.},
Observation of switchable polar skyrmion bubbles down to the atomic layers in van der Waals ferroelectric CuInP$_2$S$_6$,
\href{https://doi.org/10.1038/s41467-025-57714-9}{Nat.\ Commun.\ \textbf{16}, 2349 (2025)}.

\end{thebibliography}
\end{document}